\documentclass[a4paper,fleqn,usenatbib]{mnras}

\usepackage{txfonts}
\usepackage{multirow}
\usepackage{colortbl}
\usepackage{pdflscape}
\usepackage[T1]{fontenc}
\usepackage{ae,aecompl}

\usepackage{graphicx}	% Including figure files
\usepackage{amssymb}	% Extra maths symbols

\title[IR asteroid composition using meteorite spectra]{Role of spectral resolution for infrared asteroid compositional analysis using meteorite spectra}

\author[A. Skulteti et al.]{
A. Skulteti,$^{1}$\thanks{E-mail: skulteti.agnes@csfk.mta.hu}
A. Kereszturi,$^{2}$
Zs. Kereszty,$^{3}$
B. Pal,$^{2,5}$
M. Szabo,$^{4}$
F. Cipriani$^{6}$
\\
$^{1}$Research Centre for Astronomy and Earth Sciences, Geographical Institute, MTA Centre of Excellence, Hungary\\
$^{2}$Research Centre for Astronomy and Earth Sciences, Konkoly Thege Miklos Astronomical Institute, MTA Centre of Excellence, Hungary\\
$^{3}$International Meteorite Collectors Association\\
$^{4}$Research Centre for Astronomy and Earth Sciences, Institute for Geological and Geochemical Research, MTA Centre of Excellence, Hungary\\
$^{5}$E\"otv\"os Lor\'and University, Budapest, Hungary\\
$^{6}$European Space Agency, ESTEC/TEC-EPS, Keplerlaan 1, 2200AG, Noordwijk
}

\date{Accepted XXX. Received YYY; in original form ZZZ}

\pubyear{2020}

\begin{document}
\label{firstpage}
\pagerange{\pageref{firstpage}--\pageref{lastpage}}
\maketitle

% Abstract of the paper
\begin{abstract}
In this work the potential mineral identification of meteorites is analysed for the mid-infrared range, to evaluate observational possibilities for future missions targeting small body surfaces. Three carbonaceous and three ordinary chondrite meteorites are examined by a diffuse reflection (DRIFT) instrument, and the presence of principal minerals is confirmed by a powder diffraction method as well. The possibilities and constraints of mineral identifications in the mid-infrared are simulated by artificially degrading the spectral resolution. Our research shows that for the definite identification of principal mineral bands, a spectral resolution $\leq 10$~cm$^{-1}$ ($\leq 0.15 \mu$m) is needed. At 20-100~cm$^{-1}$ (0.3 - 1.5~$\mu$m) resolution the identification of these minerals is uncertain, and with a resolution $>100$~cm$^{-1}$, it is almost impossible.
\end{abstract}

% Select between one and six entries from the list of approved keywords.
% Don't make up new ones.

%%%%%%%%%%%%%%%%%%%%%%%%%%%%%%%%%%%%%%%%%%%%%%%%%%

%%%%%%%%%%%%%%%%% BODY OF PAPER %%%%%%%%%%%%%%%%%%

\section{Introduction}
\label{sec:intro}

The compositional analysis of asteroids is becoming more important, not only to reconstruct better the formation processes in the Solar System, but also to advance the understanding of possible methods and consequences of mitigating the threat of potentially hazardous Near Earth Asteroids \citep{rodriguez2017}, with meteorites providing direct information about them \citep{toth2011}, \citep{madiedo2013, bland1996, przylibski2003, przylibski2005}. More recently, the search for the parent bodies of possible meteorites \citep{rodriguez2007, gayon2012} has widened, partly focusing on spectral matching between asteroids and meteorites \citep{bowey2007, rodriguez2013}, which could provide information on asteroid regoliths \citep{hamm2019}.  Several missions were launched towards such asteroids in the last few years, including Hayabusa-1 \citep{yano2006} and Osiris-Rex \citep{mcmahon2018, seabrook2019}, Hayabusa-2 \citep{castelvecchi2018, michikami2019, kitazato2019, hamm2019b}. Further afield, Dawn has targeted two large main belt asteroids \citep{park2016, voosen2018}. Beyond the scientific aims, resource utilization is being considered for future missions, because some components of Near Earth Asteroids might be exploitable \citep{luszczek2019}.\\

More missions have been proposed to target Near Earth asteroids recently with cubesats \citep{kohut2018} accompanying them, for example the DART \citep{raducan2019, maindl2019} and HERA missions \citep{michel2019, tsiganis2019, carnelli2019, kueppers2019}. The use of cubesats is expected to increase in the next decades \citep{andrews2019, cheng2019, shkolnik2018} along with the improved miniaturization of various electronics and detectors. Therefore, it is worth to consider and evaluate the mineral analysing capabilities of simple and cheap infrared detectors \citep{holland2018, ardila2017}, which will probably fly regularly on such small probes. \\

The results of this research could be applied to other Solar System bodies as well, for example Phobos and Deimos, the two moons of Mars, targeted by the MMX \citep{campagnola2018, damore2019} mission planned by JAXA. The origin of these two satellites is still debated, whether they are captured asteroids \citep{burns1978,hartmann1990} or formed in place \citep{craddock2011}. Their roughly circular orbits near the equatorial plane of Mars suggest {\it in-situ} formation, however, their visible and near-infrared spectra imply the capture theory. The spectra of Phobos and Deimos show similarities to D- or T-type asteroids or carbonaceous chondrites \citep{fraeman2014, miyamoto2018}, and their spectral characteristics are mostly similar to those of Tagish Lake and CM2 chondrites. Thus, our research could be useful in the design of future missions targeting these bodies. \\

This work aims to better understand the observational possibilities of the middle infrared spectra based mineral identification \citep{martin2019, maclennan2018}, as this range provides more and detailed information on the composition than the more exploited near infrared range \citep{fulvio2018}. Using meteorite samples \citep{gilmour2019}, the almost intact interior of asteroids could be analysed \citep{donaldson2019}. Nonetheless, in reality due to space weathering produced modifications \citep{fiege2019, loeffler2018, penttil2014}, even future close-by observations might not be able to identify minerals as evidently as presented here. The laboratory based references provide important experiences for the design of middle infrared detectors, especially their channel arrangements. 

\section{Methods}
\label{sec:methods}

In this work, the DRIFT measurements based infrared spectra of some meteorites were analysed and compared using a Vertex FTIR 70 infrared spectrometer plus a Harrick based DRIFT unit called Praying Mantis,
used at room temperature. The spectral resolution is 0.04~$\mu$m (4~cm$^{-1}$), with 256 scans covering a wavelength range of 2.5-25$~\mu$m (4000-400 cm$^{-1}$). Before the measurements WERE MADE, the meteorite powders were ground down to grain sizes <50 µm, with the majority of the grains in the 10-20~µm range. The materials were dried by baking them at +120 °C for 12 hours in an oven prior to the measurements. The meteorites were ordinary chondrites from NWA 869 (L4-6 S3 W1), NWA 5838 (H6 W1), NWA 6059 (L6 S2 W2/3) and carbonaceous chondrites from Allende (CV3), NWA 10580 (CO3, S2/3, W3), NWA 11469 (CV3 type S2/3 W3). \\

A Fritsch Pulverisette-23 Mini-Ball Mill with zirconium-oxide mortar and three zirconium-oxide ball sets was used to pulverise the samples. The instrument has a 10~ml volume zirconium-oxide mortar, and works with an adjustable frequency between 15 and 50~Hz, together with adjustable pulverising duration. \\

The identified goethite was probably formed from previously existing FeNi alloy in the meteorites during weathering on the Earth. In an ideal case, only non-weathered (W=0) meteorites should be analysed, however such meteorites are difficult to obtain.  The infrared analysis here is on non-ideal, but accessible meteorite samples, providing useful results for the identification of various non-weathered silicate components. Further studies should focus on unweathered samples to identify possible differences between them and those analysed in this work. \\

The reflectance spectra were interpreted based on the spectral database of \citet{liese1975}. Spectral coarsening was done by artificially degrading the resolution with simple mathematical averaging (eq. \ref{eq:avg}, eq. \ref{eq:avgb}):
\begin{equation}
    x_d = \frac{\sum_{n=1}^{d} x}{d} 
    \label{eq:avg}
\end{equation}
\begin{equation}
      \lambda_d = \frac{\sum_{n=1}^{d} \lambda}{d}
      \label{eq:avgb}
\end{equation}
where $x_d$ represents the new data point with the desired resolution, $x$ the original data point, $d$ the resolution divisor (the degradation factor), $\lambda_d$ the wavelength of the new data point and $\lambda$ the wavelength of the original data point. After averaging, the new reduced resolution was determined by eq. \ref{eq:res}:

\begin{equation}
    \mathrm{res}_d = \lambda_d^1 - \lambda_d^0
    \label{eq:res}
\end{equation}
where $\mathrm{res}_d$ is the new resolution, and $\lambda_d^0$ and $\lambda_d^1$ are the first and the second wavelengths in the new data respectively. \\

While the detection characteristics (e.g. sensitivity along the bbservable range) of the detectors are different, a simple approach was followed here: the sensitivity was taken to be constant all along the range. This is a good enough approximation to compare the different band arrangements, as the aim was to see the general trends and general characteristics. \\

Control measurements using powder diffraction method was also applied, to have an independent estimation on the mineral composition of the analysed samples. For XRD measurements, we used a Rigaku Miniflex600 Bragg-Brentano powder-diffractometer. We milled 40 mg of the samples mixed it with 1.5 ml ethanol and dried it on a steel section to produce the XRD sample. In the Miniflex600, the X-ray source is Cu, and we used an accelerating voltage of 40 keV and a beam current of 20 mA. It uses a scintillation detector NaD graphite monochromator with a measurement time of 35 minutes. This method was applied to gain the compositonal ratio listed in Table \ref{tab:mineralcomp}.

\section{Results}
\label{sec:results}

First, the compositional analysis related general aspects are presented using the infrared band positions employed for mineral analysis, followed by results of the powder diffraction measurements. The degraded meteorite spectra are then presented, followed by a discussion of the evaluation of mineral observational possibilities.

\subsection{Mineral identification}
\label{subsec:mineralid}

The identification of minerals was done by the DRIFT spectral analysis, using the band positions in Table \ref{tab:peakpositions}. Control measurements by powder diffraction method were also made to identify the compositional ratio of minerals, shown in Table \ref{tab:mineralcomp}.\\

Although there is up to 10\% FeNi alloy in ordinary chondrites
\citep{reisener2003}, the lack of its identification might be related to two factors: (1) the possible melting and partial destruction of FeNi alloy grains during pulverization, and (2) the fact that these minerals were partially weathered to various oxides and hydroxides, including goethite.\\

\begin{table*}
	\centering
	\caption{Used infrared peak positions for mineral identification (*marks two different peaks which differ in shape also).}
	\label{tab:peakpositions}
	\begin{tabular}{lccccc} 
	    \hline
	    \multirow{2}{*}{Meteorite} & \multicolumn{5}{c}{Peak positions used for mineral identification} \\ \cline{2-6}
		& olivine & pyroxene & feldspar & troilite & goethite \\ \hline
	    Allende & 960, 842 & 669, 657 & 1092, 999 & 455 & - \\ \hline
	    NWA 5838 & 960, 842 & 1023, 669, 675 & 999 & - & 904 \\ \hline
	    NWA 10580 & 842 & 945, 911 & 1148, 1099, 999, 565 & 457 & - \\ \hline
	    NWA 6059 & 842, 506 & 945, 669, 657 & - & - & 906 \\ \hline 
	    NWA 869 & 842 & 1028 & 727, 645 & 455 & 907 \\ \hline
	    NWA 11469 & 837 & 1028 & 647, 565 & 457 & 908 \\ \hline
	\end{tabular}
\end{table*}

\begin{table*}
	\centering
	\caption{Mineral composition of meteorites used powder diffraction method.}
	\label{tab:mineralcomp}
	\begin{tabular}{lccccc} 
	    \hline
	    \multirow{2}{*}{Meteorite} & \multicolumn{5}{c}{Ratio of mineral components (\%)} \\ \cline{2-6}
		& olivine & pyroxene & feldspar & troilite & goethite \\ \hline
	    Allende & 47 & 35 & 116 & 2.155 & 0.3 \\ \hline
	    NWA 5838 & 71.6 & 24 & 2.3 & 1.5 & 0.3 \\ \hline
	    NWA 10580 & 62 & 21 & 12 & 4 & 1 \\ \hline
	    NWA 6059 & 58 & 8.3 & 1.7 & 4 & 28 \\ \hline 
	    NWA 869 & 41.5 & 25 & 26.2 & 6.9 & 0.6 \\ \hline
	    NWA 11469 & 26.8 & 21 & 28 & 0.3 & 23.9 \\ \hline
	\end{tabular}
\end{table*}

\begin{figure}
    \includegraphics[width=\columnwidth]{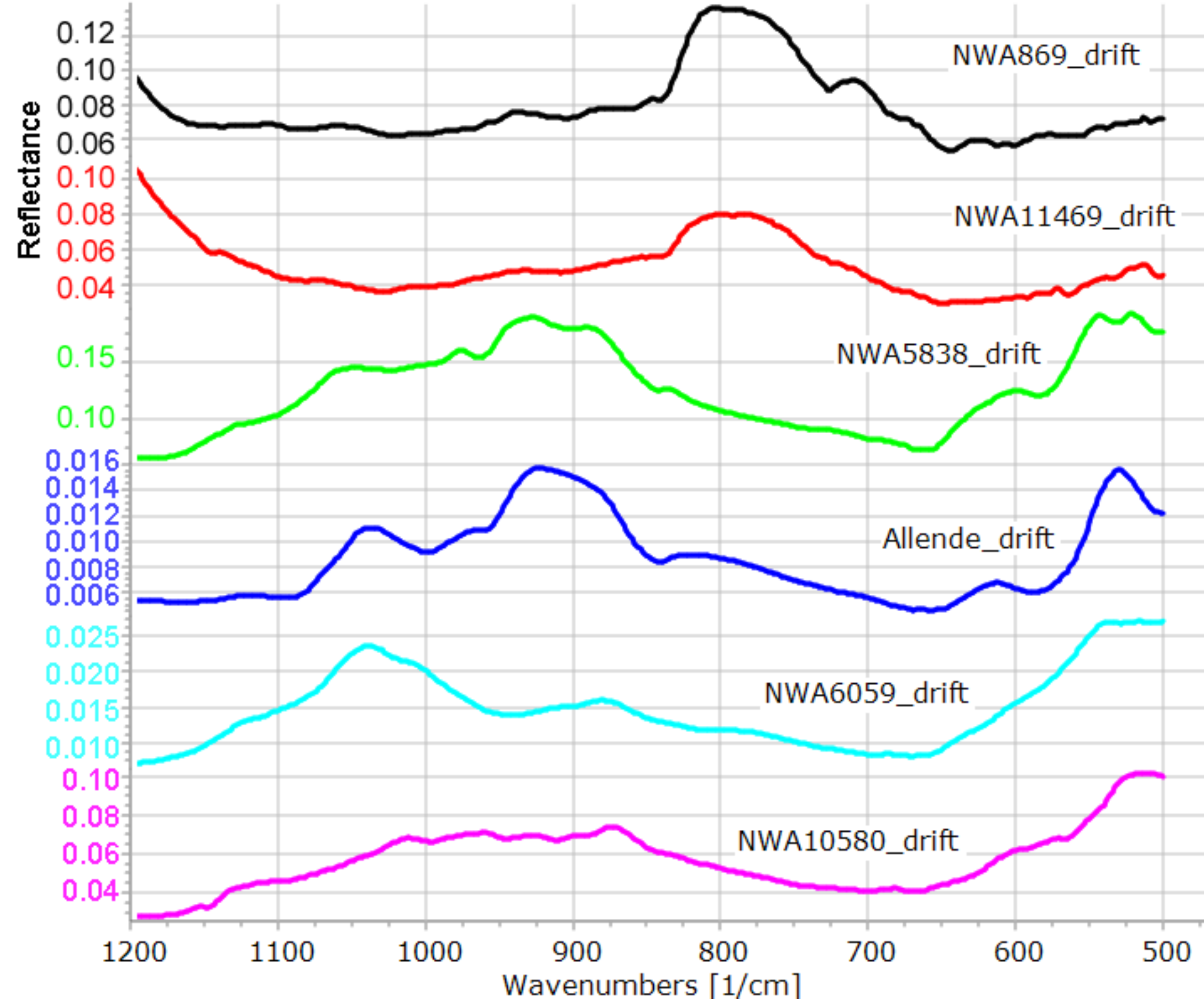}
    \caption{Mid-infrared reflectance curves of the analysed meteorites.}
    \label{fig:6gorbe}
\end{figure}

These meteorites contain several minerals with different grain sizes such that the superposition of individual peaks dominates the spectral shape, complicating interpretation. \\

The most important features of the reflectance spectra of silicates are the Christiansen and Restrahlen features. The wavelength of the Christiansen feature for individual minerals depends on polymerization (feldspar-shorter, mafic minerals: olivine, pyroxene longer wavelengths \citep{logan1973}. The Christiansen feature of pyroxene minerals is at a much shorter wavelength (8.5-8.4~µm) ($\sim$1180~cm$^{-1}$) than for olivine \citep{salisbury1991}. The Christiansen feature of feldspar minerals is at a shorter wavelength (near 8 µm) ($\sim$1250~cm$^{-1}$) than for olivine and pyroxene. The Restrahlen bands of olivine are located at 9-11.5~µm (1100-870~cm$^{-1}$) \citep{mustard1997}, while its transparency band is located at approximately 12-15~µm (830-660~cm$^{-1}$).  

\subsection{Spectral coarsening}
\label{subsec:spectralcoarse}

The observed spectra were artificially degraded to determine the minimum resolution required for the identification of key spectral features. \\

\begin{figure*}
    \includegraphics[width=17.5cm]{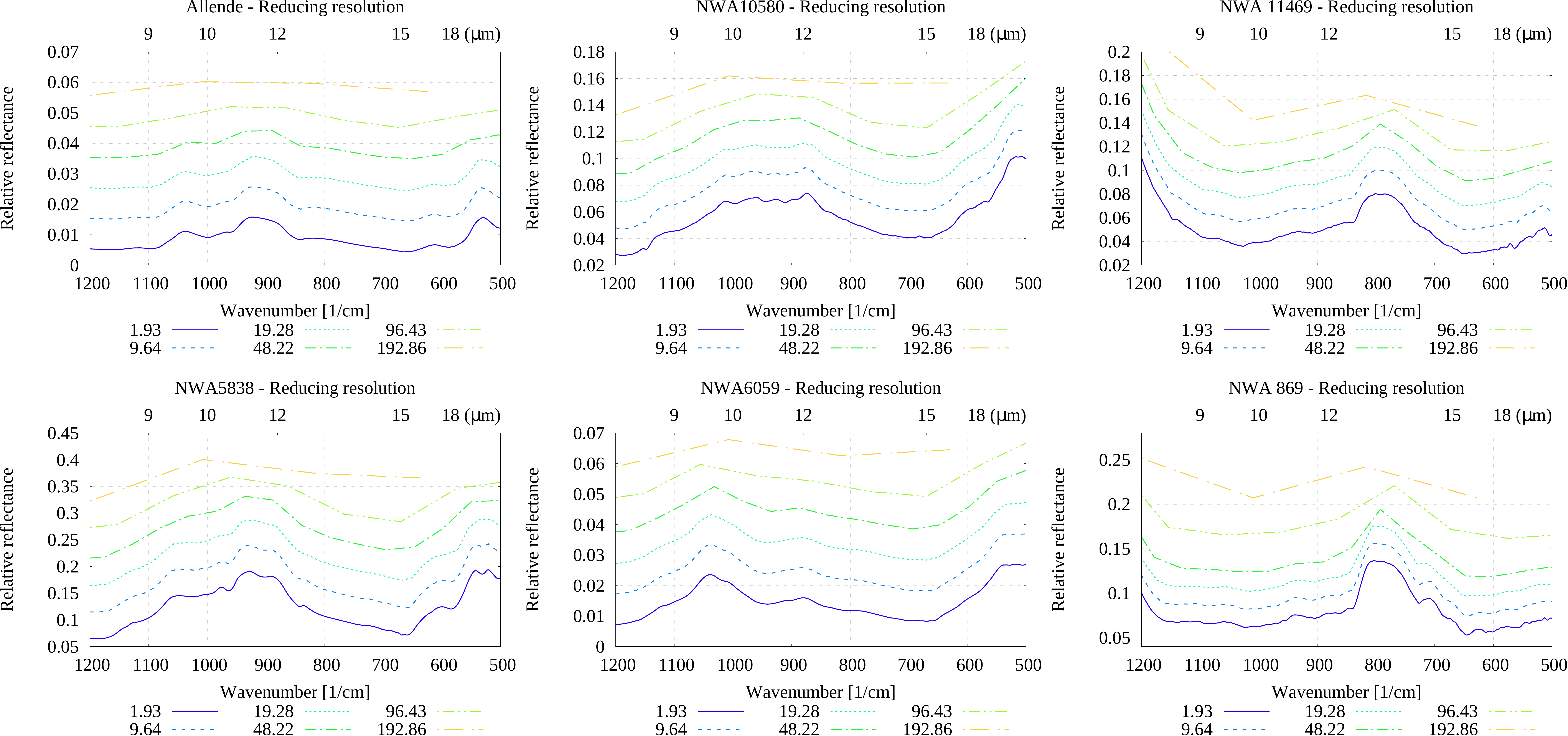}
    \caption{Results of artificial spectral degradation of the 6 analysed
      meteorites. In each diagram highest resolution can be
      found at the bottom, and gradually lower resolution spectra by the indicated colours and line styles.}
    \label{fig:spectra}
\end{figure*}

The main observable features are the common silicate mineral (olivine, pyroxene, and plagioclase) absorption bands related to iron ions. The fine details gradually disappear and the ability to identify the minerals gets worse with lower resolution. The disappearance of small peaks and fine details is clear, although the overall shape of the curves persists, even with poor spectral resolution. For example, the reflectance maxima around 800 or 1000~cm$^{-1}$ remains observable the longest,  even at the lowest resolution in some cases. 

\section{Discussion}
\label{sec:discuss}

The mineral observability is evaluated according to two different approaches: (A) The identification of more general spectral features, and (B) the identification of smaller specific mineral peaks. For the general appearance, the total shape of the curves and the following three general features were evaluated. \\ 

1. The Christiansen feature (CF, a reflectance minimum), due to a transition between the surface and volume scattering regime \citep{hapke1996}, appears as a reflectance minimum at $\sim$1100-1300~cm$^{-1}$ (7.5-9~µm) in the case of silicates. In this wavelength region the refractive index changes rapidly, which causes minimal scattering. Thus, its exact peak positions could shift and possibly overlap with other features.\\ 

2. The Restrahlen bands (local maximum in reflectance) arise from fundamental molecular vibrations as most incident radiation does not enter the sample, but is reflected on the first surface. These peaks in the reflectance at $\sim$1120-850~cm$^{-1}$ (8.5-12~µm) are by Si-O asymmetric stretching (however a less intense Restrahlen band appears around 750-300~cm$^{-1}$ (16.5-25 µm) by Si-O-Si bending).\\

3. The transparency feature (reflectance maximum) is located between two Restrahlen bands in the spectra of particulate silicates as a broad reflectance maximum, and is influenced by the presence of small particles (<75 µm) and changing optical constants. These features are diagnostic of the given minerals (e.g. \citet{logan1973, salisbury1991}). \\

Comparing the observed spectral shapes, the evaluation of the Christiansen feature might be difficult due to its complex and changing appearance, and as it also could be influenced by the Restrahlen feature, which could overlap it. A reflectance minimum around the expected Christiansen feature was present in the case of Allende, NWA 5838, NWA 6059, NWA 10580, however it emerged, but shifted to longer wavelengths between 9-12~µm. This minimum seemed to be present in the spectrum of NWA 11469, but not with certainty. These observations suggest that if one aims to identify the band (and drop) in reflection compared to about 10-12~µm, it could be lost. If one searches for the the edges of the band along different wavelengths but considers the range only at 8-9~µm, the Christiansen feature could be missed, as it might emerge at around 9-12~µm. It could also cause problems if there is a stronger reflection at the expected location, and no bands emerge relatively to the above mentioned “shoulder” location. The candidate bands for the Restrahlen feature appeared  approximately at its expected location (8-12~µm) for NWA 869 AND NWA 11469. However, for the other four meteorites (Allende, NWA 5838, NWA 6059, NWA 10580) there was no band but just an occasionally elevated reflectance.\\

With low resolution, the Christiansen, Restrahlen and transparency features are not easily observable and thus are not ideal targets for identification.  This is partly because they show variability; in fact, their detection is not straightforward even at high resolution. In the case of meteorites, The Christiansen feature is a combination of the characteristic CFs of all the constituent minerals. The Restrahlen features generally did not turn out to be a characteristic structure in the observed spectra, either. To summarize, to see the structure of Christiansen or Restrahlen bands, a spectral resolution of about 0.1~µm is required between 8-12~µm. In contrast, there were silicate rich meteorites that did not present these features, or at least not clearly. No general rule for the identification of silicate features could be found. This suggests that under non-ideal conditions, targeting “real”, space-weathered asteroid surfaces, the identification of these spectral features might be even more difficult. The variability between meteorites of the same type also suggests that any mineral identification method needs to be more general, requiring further testing. 

\subsection{Identification of specific minerals}
\label{subsec:specminid}

It is probably more feasible to identify specific mineral identifiers than to rely on general spectral features. To evaluate this possibility at different spectral resolutions, the most important peaks of principal minerals (olivine, pyroxene, felspar and goethite) were surveyed and characterized using the following grades: 2 – evident band, 1 – uncertain band, 0 – no band visible (Table \ref{tab:occurrence}). To make firm identifications of the main bands of mineral components, a spectral resolution better than 10~cm$^{-1}$ (0.15~µm) is required, with equally distributed bands in the 800-1000~cm$^{-1}$ (10-13~µm) range. The identification of these minerals is problematic and uncertain around resolution 20-100~cm$^{-1}$ (0.3-1.5 µm), and almost impossible with a resolution worse than $\sim$100~cm$^{-1}$ (1.5 µm). \\

\begin{figure*}
	\includegraphics[width=17.5cm]{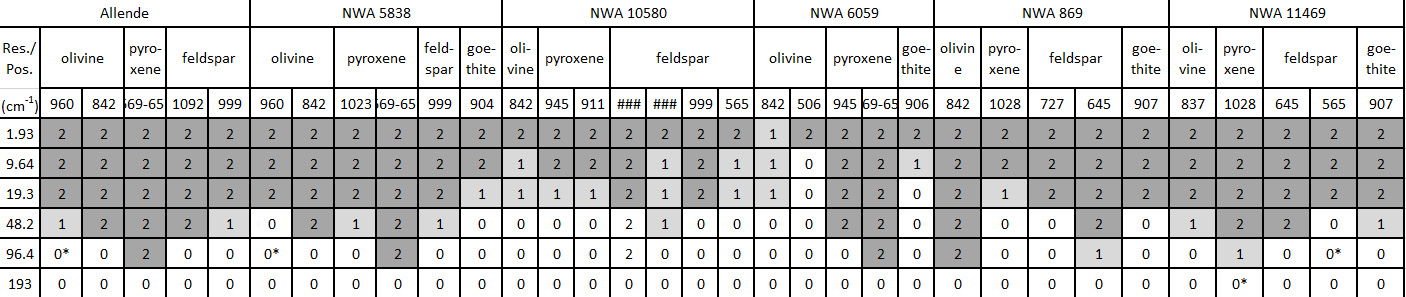}
   \caption{Overview of the observability of selected spectral band positions, characteristic for the given minerals.}
   \label{fig:observability}
\end{figure*}

Future research projects should cover lower weathering grade samples too, as here the otherwise present FeNi alloys were transformed to mainly goethite, but the results are still valid for silicates.

\subsection{Proposed channel arrangement}
\label{subsec:channel}

Although the main bands of characteristic minerals could be
identified, it is difficult to identify minerals successfully for all six meteorites using the same bands. In practice, any asteroid mission needs prepared for mineral identification using the same spectral set-up as lab-tested on related meteorites.\\

Two general problems emerge during the mineral identification.  One is that bands of the same minerals might vary slightly, thus requiring a wider filter.  However, this width is not large, as seen in Table \ref{tab:peakpositions}. In the case of olivine, one key band ranged between 842-837~cm$^{-1}$, while the other appeared in both cases at 960~cm$^{-1}$. For pyroxene, 657-669~cm$^{-1}$ and 1023-1028~cm$^{-1}$ were observer. In contrast, feldspar 999 and 565 did not change, while a shift between 1092-1099~cm$^{-1}$ was observed for a third meteorite. For troilite the band was 455-457~cm$^{-1}$, while for goethite it was 904-908~cm$^{-1}$. In summary, where the same band appeared for different meteorites, the range of its shift was $\sim$4-9~cm$^{-1}$, which might be wide enough for its identification. \\

The other difficulty was that for the same minerals different bands were observed for different meteorites e.g. a given band did not suffice for all meteorites. Table~\ref{tab:occurrence} gives the bands as a function of mineral and meteorite. Peaks are listed only if they were observed for at least two different meteorites. Despite all five minerals being present in each meteorite, they could be identified by the IR reflection spectra only in about half of the cases. As the analysed samples came from pulverized and homogenized powder of the meteorites, spatial compositional inhomogeneity could not produce this observed characteristic. \\

\begin{table*}
	\centering
	\caption{Occurrence of specific mineral bands on the analysed
          meteorites in units of cm$^{-1}$. Only those
          bands are listed where they are observed in at least two meteorites.}
	\label{tab:occurrence}
	\begin{tabular}{l>{\columncolor[gray]{0.8}}c>{\columncolor[gray]{0.8}}ccccc>{\columncolor[gray]{0.8}}c>{\columncolor[gray]{0.8}}c>{\columncolor[gray]{0.8}}ccc} 
	    \hline
	    \multirow{2}{*}{Meteorite} & \multicolumn{2}{c}{Olivine} & \multicolumn{4}{c}{Pyroxene} & \multicolumn{3}{c}{Feldspar} & Troilite  & Goethite \\ \cline{2-12}
		& 842 & 960 & 657 & 669 & 945 & 1023 & 565 & 999 & 1092 & 455 & 904 \\ \hline
	    Allende & X & X & X & X & & & X & X & X & & \\ \hline
	    NWA 869 & & & & & & X & & & X & X \\ \hline
	    NWA 5838 & X & X & X & X & & X & & X & & X \\ \hline
	    NWA 6059 & & & X & X & X & & & & & & X \\ \hline 
	    NWA 10580 & X & & & & X & X & X & X & X & &  \\ \hline
	    NWA 11469 & & & & & & X & X & & & X & X \\ \hline
	\end{tabular}
\end{table*}

In agreement with the expectations, the concentration of minerals does not correlate with their dominance (and occasionally even their presence) in the spectra. Consequently, similar to laboratory results, the identification of given peaks does not correspond to concentration.  However, a large or small areal occurrence at an asteroid surface might have a weak correlation with abundance (giving a rough upper limit for global occurrence). Based on theoretical considerations, crystalline structure (purity, homogeneity, and amorphous component ratio), grain size and possibly superposing spectral features from other minerals might cause the above identified diversity in the spectral identification of minerals. \\

Although the overall spectral shapes differed between the meteorites, NWA 869 (OC) and 11469 (CC) had some spectral similarities, as did NWA 5838 (OC) and Allende (CC). Considering their different types, there appears to be no characteristic trend. However, this may be expected as their main mineral components were the same. Additionally, the carbonaceous component probably does not have a major impact on the general spectral shape.

\subsection{Instrumental experiences}
\label{subsec:instrumental}

Table~\ref{tab:detectors} lists the basic spectral parameters for recently launched or planned asteroid missions. Most detectors would be able to identify the principal minerals in this study, although appropriate filters or selection of wavelengths would be necessary. Given the result that in roughly half of the cases a given mineral could be identified by multiple band features, assuming 3 filters for a given band identification, the identification of 3-4 minerals seems possible using 9-12 filters.\\

\begin{table*}
	\centering
	\caption{Characteristics of mid-infrared detectors of recently launched and planned missions.}
	\label{tab:detectors}
	\begin{tabular}{p{4cm}p{5cm}p{1.5cm}p{2cm}p{3cm}} 
	    \hline
	    Mission name (target object) & Instrument name & Covered range & Spectral \newline resolution & References and/or \newline example results \\ \hline
	    BepiColombo (Mercury) & MERTIS (MErcury Radiometer and Thermal Infrared Spectrometer) & 7-14 µm & 0.09 µm & \citet{hiesinger2018} \\ \hline
	    MarcoPolo-R (asteroid, cancelled) & THERMAP (THERMalMApper) & 8-16 µm & 0.3 µm & \citet{groussin2016} \\ \hline 
	    Osiris-Rex (Bennu) & OTES (Osiris-Rex Thermal Emission \newline Spectrometer) & 4-50 µm & 0.15 µm & \citet{christensen2018} \\ \hline
	    Hayabusa-2 (Ryugu) & Thermal Infrared Imager (TIR) & 8-12 µm & 0.02 µm & \citet{okada2018} \\ \hline 
	    Hayabusa-2 MASCOT lander \newline  (Ryugu) & MicrOmega & 0.99-3.65 µm & 0.3 µm & \citet{bibring2017} \\ \hline 
	    Lunar Reconnaissance Orbiter \newline (Moon) & Diviner Lunar Radiometer Experiment & 0.3-400 µm & 9 channels & \citet{chin2007} \\ \hline 
	    AIM, HERA mission \newline (Didymos asteroid) & TIRI & MIR & 7-10 channels & \citet{manzillo2018} \\ \hline
	\end{tabular}
\end{table*}

\section{Conclusion}
\label{sec:conclusion}

In this work the potential mineral identification of meteorites in the mid-infrared range as a function of spectral resolution was studied, in order to evaluate similar observational possibilities for asteroid surfaces by future space missions. Altogether 3 carbonaceous and 3 ordinary chondrite meteorites were investigated by a diffuse reflection providing DRIFT instrument. The existence of principal minerals was confirmed by a power diffraction method as a comparison to the infrared analysis. \\

Spectral bands covering 800-1000~cm$^{-1}$ (10-13~µm) were used to infer the mineral composition. Three potential issues were evaluated using the observed infrared reflectances. As expected, the dominance and strength of certain peaks of a given mineral do not correlate with its abundance/concentration in the sample. The same mineral might produce different peaks in different samples, making it hard to specifically identify its presence using a few predefined band positions. As a result, during the detector design phase, more peaks for one mineral should be considered. The most useful positions based on the laboratory survey of this work are the following: olivine: 842, 960~cm$^{-1}$; pyroxene 657, 669, 945, 1023~cm$^{-1}$; feldspar: 565, 999, 1092~cm$^{-1}$; troilite: 455~cm$^{-1}$; and goethite: 904~cm$^{-1}$. \\

The identification of principal minerals at different resolutions (by artificially degrading the spectra) was compared.  For definite identification of the main mineral bands, resolution of $<$10~cm$^{-1}$ (0.15 µm), with equally distributed bands. Around 20-100~cm$^{-1}$ (0.3-1.5 µm) resolution the identification is uncertain, and it is not possible with a resolution $>$100~cm$^{-1}$ (1.5 µm). \\

Mission planners should be prepared to design mid-infrared spectrographs to cover multiple band features for a given mineral. Based on a sample of six carbonaceous and ordinary chondrite meteorites, the most important ones have been identified.  These findings might have special importance in the case of cubesats, as this probe class is getting involved more frequently in space missions. On-board cubesats' low cost infrared detectors could provide optimised data, if the filter or band positions are carefully selected. Results of this work provide a basis for the refinement of future spectral band selection.

\section*{Acknowledgements}

This work was supported by the NEOMETLAB project of ESA for the
spectral data, by the Excellence of Strategic R\&D centres
(GINOP-2.3.2-15-2016-00003) project of NKFIH and the related H2020 fund FOR the meteorite sample access, by the GINOP 2.3.2-15-2016-00009 for the mineral composition of meteorites, and by the the ÚNKP-19-3 New National Excellence Program of the Ministry for Innovation and Technology.  \\

%%%%%%%%%%%%%%%%%%%%%%%%%%%%%%%%%%%%%%%%%%%%%%%%%%

%%%%%%%%%%%%%%%%%%%% REFERENCES %%%%%%%%%%%%%%%%%%

% The best way to enter references is to use BibTeX:

\bibliographystyle{mnras}
\bibliography{IRAsteroidComp_Review2_CLEAN} % if your bibtex file is called example.bib

% Alternatively you could enter them by hand, like this:
% This method is tedious and prone to error if you have lots of references
%\begin{thebibliography}{99}
%\bibitem[\protect\citeauthoryear{Author}{2012}]{Author2012}
%Author A.~N., 2013, Journal of Improbable Astronomy, 1, 1
%\bibitem[\protect\citeauthoryear{Others}{2013}]{Others2013}
%Others S., 2012, Journal of Interesting Stuff, 17, 198
%\end{thebibliography}

%%%%%%%%%%%%%%%%%%%%%%%%%%%%%%%%%%%%%%%%%%%%%%%%%%

%%%%%%%%%%%%%%%%% APPENDICES %%%%%%%%%%%%%%%%%%%%%

%\appendix

%\section{Some extra material}

%If you want to present additional material which would interrupt the flow of the main paper,
%it can be placed in an Appendix which appears after the list of references.

%%%%%%%%%%%%%%%%%%%%%%%%%%%%%%%%%%%%%%%%%%%%%%%%%%

% Don't change these lines
\bsp	% typesetting comment
\label{lastpage}
\end{document}